\documentclass[runningheads]{llncs}
\usepackage{graphicx}
\begin{document}
\title{Rusty Links in Local Chains\thanks{Supported by the Royal Society of New Zealand Marsden Grant, and Agoric Inc..}}
%
%
\author{James Noble\inst{1}\orcidID{0000-0001-9036-5692} \and
Julian Mackay\inst{2}\orcidID{0000-0003-3098-3901} \and
Tobias Wrigstad\inst{3}\orcidID{0000-0002-4269-5408}}
\authorrunning{J.\ Noble, J.\ Mackay et al.}
%
\institute{
Creative Research \& Programming, Darkest Karori, 
\email{kjx@acm.org}\and 
Julian Mackay, Victoria University of Wellington, \email{Julian.Mackay@ecs.vuw.ac.nz} \and
Tobias Wrigstad, Uppsala University, 
\email{tobias.wrigstad@it.uu.se}}
\maketitle              
\begin{abstract}
Rust successfully applies ownership types to control memory allocation. This restricts the programs' topologies to the point where doubly-linked lists cannot be programmed in Safe Rust. We sketch how more flexible ``local'' ownership could be added to Rust, permitting multiple mutable references to objects, provided each reference is bounded by the object's lifetime. To maintain thread-safety, locally owned objects must remain thread-local; to maintain memory safety, local objects can be deallocated when their owner's lifetime expires.

\end{abstract}
\section{Rusty Links}

Rust \cite{RustBook,RustPopular,MSRust} is well-known as a language that combines control of memory use, safe concurrency, and excellent compiler error messages.  Rust achieves this balance thanks to a version of \textit{ownership types} \cite{ClaPotNobOOPSLA98,NobPotVitECOOP98,OTSurvey}
(also known in the literature as ``\textit{ownership types}''
\cite{SafeRust,djpRust}) which statically track the lifetime (or owner) of each allocated object; when an object goes out of scope, all the memory owned by that object is deallocated. So far, so C$++$ \cite{cpp}, but Rust's ownership types ensure that programs remain memory safe, so really not C$++$. 
Rust then incorporates borrowing \cite{borrowing} and fractional permissions \cite{boyland:2003:fractional} to support an integral multiple-reader/single-writer concurrency model \cite{lea98}: at any time, an object may either be accessed by multiple read-only aliases, or by a single read-write reference.

Rust's ownership types are necessarily conservative, banning not just all concurrent programs that are \textit{actually} unsafe, but a large number of data-race free programs as well.  To programmers, this
manifests as a large number of \textit{false positive} errors or
warnings about problems that will never arise in practice. Many programmers find  Rust hard to learn and to use correctly \cite{LearnRust,VizRust,HardRust,SafeRust,FightRust}.  
%
Rust's version of ownership types \cite{RustBook} bans common idioms such as circular or doubly-linked lists,
to the point where the difficulty of implementing a data structure often taught at first year has now become an Internet trope \cite{RustDLL1,RustDLL2,RustDLL3,RustDLL4}.  
A number of solutions have been proposed for these problems, including incorporating a garbage collector \cite{rustGC}, careful library design \cite{RustMSc}, phantom types \cite{GhostCell}, or proving unsafe Rust code correct \cite{RustBelt18,StackedBorrows}.

\section{Local Chains} 

We propose to solve this problem by supporting \textit{thread local} ownership within Rust. Rust's type system currently supports two kinds of borrowing of a variable \verb+v+.
Writing ``\verb+& foo+'' gains readonly access to \verb+v+, which allows multiple aliasing; while writing \verb+&mut v+ grants read/write access to only one active alias. For example, we can establish two active readonly references to  a variable \verb+v+ but we cannot assign to either reference, even though the underlying variable is mutable:

\begin{verbatim}
   let mut v : i32 = 12;

   let a = &v;
   let b = &v;
   println!("{:#?}", a);  //read a 
   println!("{:#?}", b);  //read b
   //*a = 45; //a is not mutable, cannot write
\end{verbatim}

\noindent Alternatively, we can establish one mutable reference to \verb+v+ through which we can change \verb+v+'s value:

\begin{verbatim}
   let c = &mut v;
   //let d = &mut v; //cannot borrow `v` as mutable more than once
   *c = 45;
   println!("{:#?}", c);
\end{verbatim}

We propose to add a third kind of borrowing --- local ownership --- which permits both aliases and mutability. We can establish multiple local references to \verb+v+ by writing ``\verb+&loc v+'' and can change \verb+v+'s value through all of them:
\begin{verbatim}
   let loc v : i32 = 12;

   let e = &loc v;
   let f = &loc v; // two local read/write borrows
   *e = 67;
   *f = 76;
   println!("{:#?}", e);   println!("{:#?}", f);
\end{verbatim}

\noindent
These local aliases should be enough to support chains of mutable objects.
To be safe, local objects can only be accessed locally: they cannot be shared or moved, and must remain  within one thread. Rust's ownership deallocates objects  whenever they go out of scope. Because local objects can be internally aliased, 
we cannot deallocate them individually: rather we must deallocate all the local objects in one operation at the end of their \textit{owner's} scope.  We can explore 
per scope memory allocation patterns:  fixed size and extensible arenas, reference counting, and
even garbage collection, as e.g.\ in Real-Time Java \cite{smsbook,rtsjmem},  with  extensions to finer-grained scopes, alias analysis, and safe \emph{manual} memory management.

Finally,  we hope this approach could inform (and be informed by) formal techniques for other "Rust-like"  languages such as 
Pony \cite{PonyTS},
Encore \cite{EncoreTS},
Deterministic Parallel Java \cite{DPJ},
Obsidian \cite{aldrichObsidianStudy2020}, 
Dala \cite{Dala}, and
Verona \cite{Verona}.

%
%
%
\bibliographystyle{splncs04}
\bibliography{Case}

\begin{thebibliography}{10}
\providecommand{\url}[1]{\texttt{#1}}
\providecommand{\urlprefix}{URL }
\providecommand{\doi}[1]{https://doi.org/#1}

\bibitem{LearnRust}
Abtahi, P., Dietz, G.: Learning {R}ust: How experienced programmers leverage
  resources to learn a new programming language. In: {CHI} Extended Abstracts.
  pp.~1--8 (2020)

\bibitem{RustMSc}
Beingessner, A.: You can't spell Trust without {R}ust. Master's thesis,
  Computer Science, Carleton University (2015)

\bibitem{RustDLL1}
Beingessner, A.: Learn {R}ust with entirely too many linked lists.
  \texttt{https://\-rust-unofficial.github.io/\-too-many-lists} (Mar 2019),
  accessed April Fools Day 2022

\bibitem{VizRust}
Blaser, D.: Simple explanation of complex lifetime errors in {R}ust (2019),
  {ETH Z{\"u}rich}

\bibitem{DPJ}
Bocchino, R., Heumann, S., Honarmand, N., Adve, S., Adve, V., Welc, A.,
  Shpeisman, T.: {Safe Nondeterminism in a Deterministic-by-Default Parallel
  Language}. In: POPL (2011)

\bibitem{rtsjmem}
Bollella, G., Canham, T., Carson, V., Champlin, V., Dvorak, D.L., Giovannoni,
  B., Indictor, M.B., Meyer, K., Murray, A., Reinholtz, K.: Programming with
  non-heap memory in the real time specification for {J}ava. In: {OOPSLA}
  Companion. pp. 361--369 (2003)

\bibitem{borrowing}
Boyland, J.: Alias burying: Unique variables without destructive reads.
  Software: Practice \& Experience  \textbf{31}(6) (May 2001)

\bibitem{boyland:2003:fractional}
Boyland, J.: Checking interference with fractional permissions. In: Static
  Analysis Symposium. pp. 55--72 (2003)

\bibitem{RustDLL4}
Cameron, N.: What’s the “best” way to implement a doubly-linked list in
  {R}ust?
  \texttt{http://featherweightmusings.blogspot.com/2015/04/graphs-in-rust.html}
  (Apr 2015), accessed April Fools Day 2022

\bibitem{EncoreTS}
Castegren, E., \textrm{Tobias Wrigstad}: Reference capabilities for concurrency
  control. In: ECOOP (2016)

\bibitem{Verona}
Chisnall, D., Parkinson, M., Clebsch, S.: Project {V}erona (2021),
  \texttt{www.microsoft.com/\-en-us/\-research/\-project/\-project-verona}

\bibitem{OTSurvey}
Clarke, D., {\"{O}}stlund, J., Sergey, I., \textrm{Tobias Wrigstad}: Ownership
  types: {A} survey. In: Aliasing in Object-Oriented Programming. Types,
  Analysis and Verification, {LNCS}, vol.~7850 (2013)

\bibitem{ClaPotNobOOPSLA98}
Clarke, D., Potter, J.M., \textrm{James Noble}: Ownership types for flexible
  alias protection. In: OOPSLA (1998)

\bibitem{PonyTS}
Clebsch, S., et~al.: Deny capabilities for safe, fast actors. In: {AGERE}. pp.
  1--12 (2015)

\bibitem{rustGC}
Coblenz, M., Mazurek, M.L., Hicks, M.: Does the bronze garbage collector make
  {R}ust easier to use? {A} controlled experiment. In: {ICSE} (2022)

\bibitem{aldrichObsidianStudy2020}
Coblenz, M.J., Aldrich, J., Myers, B.A., Sunshine, J.: Can advanced type
  systems be usable? an empirical study of ownership, assets, and typestate in
  {O}bsidian. {OOPSLA}  (2020)

\bibitem{RustDLL3}
Cohen, R.: Why writing a linked list in (safe) {R}ust is so damned hard.
  \texttt{https://rcoh.me/posts/rust-linked-list-basic\-ally-impossible/} (Feb
  2018), accessed April Fools Day 2022

\bibitem{Dala}
Fernandez{-}Reyes, K., Gariano, I.O., \textrm{James Noble},
  Greenwood{-}Thessman, E., Homer, M., \textrm{Tobias Wrigstad}: Dala: A simple
  capability-based dynamic language design for data race-freedom. In: {Onward!}
  (2021)

\bibitem{RustPopular}
Hu, V.: Rust breaks into {TIOBE} top 20 most popular programming languages (Jun
  2020), infoQ

\bibitem{StackedBorrows}
Jung, R., Dang, H.H., Kang, J., Dreyer, D.: Stacked borrows: An aliasing model
  for {R}ust. In: {POPL} (2019)

\bibitem{RustBelt18}
Jung, R., Jourdan, J.H., Krebbers, R., Dreyer, D.: Rustbelt: Securing the
  foundations of the rust programming language. {PACMPL}  \textbf{2}(POPL),
  66:1--66:34 (Jan 2017)

\bibitem{SafeRust}
Jung, R., Jourdan, J.H., Krebbers, R., Dreyer, D.: {Safe Systems Programming in
  Rust: The Promise and the Challenge}. {Communications of the ACM}  (2020)

\bibitem{RustBook}
Klabnik, S., Nichols, C.: The Rust Programming Language. 2nd edn. (2018)

\bibitem{MSRust}
Krill, P.: Microsoft forms {R}ust language team (Feb 2021), infoWorld

\bibitem{lea98}
Lea, D.: Concurrent Programming in {Java}. Addison-Wesley, 2nd edn. (Dec 1998)

\bibitem{RustDLL2}
ndrewxie: What’s the “best” way to implement a doubly-linked list in
  {R}ust?
  \texttt{https://users.\-rust-lang.org/\-t/\-whats-the-best-way-to-implement-a-doubly-linked-list-in-rust/\-27899/\-7}
  (Mar 2019), accessed April Fools Day 2022

\bibitem{smsbook}
Noble, J., Weir, C.: Small Memory Software: Patterns for systems with limited
  memor y. Addison-Wesley (2000)

\bibitem{djpRust}
Pearce, D.J.: A lightweight formalism for reference lifetimes and borrowing in
  {R}ust. {TOPLAS}  \textbf{43}(1) (2021)

\bibitem{HardRust}
Qin, B., Chen, Y., Yu, Z., Song, L., Zhang, Y.: Understanding memory and thread
  safety practices and issues in real-world {R}ust programs. In: {PLDI}. pp.
  763--779 (2020)

\bibitem{FightRust}
Spencer, R.J.: Four ways to avoid the wrath of the borrow checker (2020),
  justanotherdot.com

\bibitem{cpp}
Stroustrup, B.: The C++ Programming Language (1986)

\bibitem{NobPotVitECOOP98}
\textrm{James Noble}, Potter, J., Vitek, J.: Flexible alias protection. In:
  ECOOP (Jul 1998)

\bibitem{GhostCell}
Yanovski, J., Dang, H., Jung, R., Dreyer, D.: Ghostcell: separating permissions
  from data in {R}ust. In: {ICFP} (2021)

\end{thebibliography}

\end{document}